\documentclass [11pt]{article}

\usepackage{graphicx,amsmath,amssymb,latexsym,epsfig,epstopdf, cite}
\usepackage{flushend}
\usepackage{color}
\usepackage{algorithm}
\usepackage{algorithmic}
\usepackage{caption}
\usepackage{subcaption}
\usepackage{lmodern,textcomp}
\usepackage{dblfloatfix}

\def\etal{{\em et al. }}

\begin{document}

\title{Exploring the Memory-Bandwidth Tradeoff in an Information-Centric Network}

\author{
{James Roberts (Inria and SystemX, France)}
\and
{Nada Sbihi (Inria, France)}
}

\date{}

\maketitle

\begin{abstract}
An information-centric network should realize significant economies by exploiting a favourable memory-bandwidth tradeoff: it is cheaper to store copies of popular content close to users than to fetch them repeatedly over the Internet. We evaluate this tradeoff for some simple cache network structures under realistic assumptions concerning the size of the content catalogue and its popularity distribution. Derived cost formulas reveal the relative impact of various cost, traffic and capacity parameters, allowing an appraisal of possible future network architectures. Our results suggest it probably makes more sense to envisage the future Internet as a loosely interconnected set of local data centers than a network like today's with routers augmented by limited capacity content stores.
\end{abstract}


\section{Introduction}
It has become a commonplace to observe that the Internet has become information-centric, with more than 90\% of its traffic resulting from content retrieval\footnote{See Cisco Visual Network Index, for example.}. There is broad agreement that the network should make more extensive use of caching in order to exploit an increasingly favourable memory-bandwidth tradeoff: it is cheaper to store copies of content items close to users than to repeatedly fetch them from some remote server. This tradeoff determines the optimal cache size, number and location and therefore has a strong impact on the structure of the future Internet. Our objective in this paper is to derive quantitative tradeoff results that reveal the structurally determinant parameters of an information-centric network (ICN).

Most currently proposed novel ICN architectures would systematically equip routers with caches in order to reduce the volume of content-retrieval traffic \cite{ADIKO12}. However, the effectiveness of such universal caching has recently been called into question \cite{Ghodsi2011}. We agree with the authors of this paper that the in-network caching assumption needs closer analysis, though not for the same reasons. The arguments in  \cite{Ghodsi2011} are based on models where cache size is assumed infinite, effectively supposing storage costs are negligible. Our own doubts stem rather from the observation formulated in \cite{FRRS12} that, to be effective in significantly reducing traffic volumes, cache sizes need to be very large, orders of magnitude larger than the storage that could reasonably be added to a router  \cite{PV11}. Rather than a network of content routers equipped with caches, the future Internet might more realistically be seen as a network of data centers that, among other applications, also do routing. 

The memory-bandwidth tradeoff depends on the cache hit rate that determines the proportion of download traffic that is saved by a cache of given size. We use the ``Che approximation'' to evaluate the hit rate of a cache assuming least recently used (LRU) replacement under the independent reference model (IRM) \cite{CTW02,FRR12}. The hit rate depends crucially on the size and popularity distribution of the considered content catalogue. To derive a realistic characterization we have used data recorded by Dan and Carlsson \cite{DC10} for content retrieved using BitTorrent. These data allow us to directly derive the relative torrent request rates, as required for the IRM. In the absence of comparable measurement results for other types of content, we use this as a generic popularity law with characteristic head, body and tail behaviour. 
 
To more clearly identify structural properties we consider simple symmetrical cache networks. The base case is a simple 2-level hierarchy where users first address requests to their local level-1 cache. In case of a miss, requests are redirected to a single central cache at level-2. The tradeoff is determined by the total cost of storage compared to the cost of the bandwidth needed to handle peak traffic flowing between the two levels. We use simple cost formulas that allow straightforward understanding of the impact of key parameters. This is important since assumed parameter values are necessarily imprecise and subject to quite rapid change as costs tend to decrease while demand and content catalogues grow. 

Results for the simple 2-level hierarchy highlight structural issues but costs might be further reduced by using some form of inter-cache cooperation. We therefore extend the tradeoff analysis to account for two generalizations drawn from the literature. First, we suppose level-1 caches perform load sharing by specializing the content they store. A hash of the chunk name determines to which level-1 cache the request must be sent. In the second generalization, users send all requests to their local cache but in case of a miss, a request is sent to another level-1 cache that, as for load sharing, is designated by a hash of the chunk name.  
These schemes require less storage than the basic hierarchy but incur additional bandwidth costs leading to a modified tradeoff. 

The rest of the paper is organized as follows. We begin in the next section by situating our approach with respect to related work. In Section \ref{sec:Chepop} we recall the Che approximation and use it to derive some useful properties of the model. This section also derives the BitTorrent popularity law used to evaluate the memory-bandwidth tradeoff. The memory-bandwidth tradeoff is evaluated in detail for the 2-level cache hierarchy in Section \ref{sec:hierarchy}. Generalizations of this analysis to more sophisticated cooperative cache networks are described in Section \ref{sec:alternatives}.

\section{Related work}
\label{sec:related}
There is a vast literature on how to place content items in order to optimize the memory-bandwidth tradeoff under various constraints. The papers cited below are a small sample meant to illustrate of the approaches that are most relevant to the present objective.

A paper by Nussbaumer \etal \cite{NPSS95} adopts a similar approach to ours in directly comparing the costs of storage and bandwidth. The authors envisage caches at different levels of a symmetric tree hierarchy and study cost as a function of cache size. Their results are not directly applicable, however, mainly because we have quite different cost assumptions. Cidon \etal  \cite{CKS2002 } propose a method for minimizing the sum of bandwidth and storage costs under general assumptions but provide no numerical results or qualitative analysis. 

A number of papers seek to minimize bandwidth usage when demand is assumed known for each object. Kangasharju \etal \cite{KRR01} supposes cache locations and capacities are given while Laoutaris \etal \cite{LZS05} proposes heuristics to jointly optimize storage allocation and content placement under a constraint on overall storage capacity. In both cases, the storage cost is replaced by capacity constraints and the analysis therefore brings little insight concerning the memory-bandwidth tradeoff.

Two recent papers revisit the issue of optimal content placement for video-on-demand. Borst \etal \cite{BGW10} seeks to minimize bandwidth usage under given cache capacities and known demand. This work is similar to ours in that the authors characterize the nature of the optimal solution for a simple symmetric 2-level cache hierarchy. Applegate \etal \cite{Applegate2010} optimally places video content in a given network where both storage capacity and link bandwidths are fixed. Again, since capacity is fixed \emph{a priori}, the results of these papers cannot be used to evaluate the memory-bandwidth tradeoff. 
  
In the large body of current work on a future information-centric Internet, we have found little of direct use in our evaluation. Most papers follow Jacobson \etal \cite{JSTP09} in assuming caching is performed by routers equipped with a content store. Since such a store is limited in capacity for technological reasons, the aim is to optimally exploit overall storage capacity by various selective and cooperative caching strategies (e.g., \cite{LS11, RR12, CHPP12}).  In order to avoid very poor performance under the assumed capacity constraints, it appears necessary either to assume the content catalogue is unrealistically small or that demand is unreasonably concentrated on the most popular items. In our work we do not pre-suppose in-router caching and use a realistic content retrieval model derived from measurements.

\section{Cache performance and popularity laws}
\label{sec:Chepop}

We recall the Che approximation and derive a popularity law from measurements of BitTorrent activity.
\subsection{The Che approximation}
\label{sec:Che}
Cache hit rates are derived using the approximation introduced by Che \etal \cite{CTW02} and shown by Fricker \etal \cite{FRR12} to be extremely accurate, especially for the large cache sizes considered here. 
The ``Che approximation'' uses the independent reference model (IRM) where users request objects from a fixed catalogue of size $N$, the probability a request is for some object $n$ being fixed and independent of all prior requests. Let the latter probability be proportional to a ``popularity law'' $q(n)$ for $1\le n \le N$. Under the IRM, the $q(n)$ can be interpreted as rates of independent Poisson processes. 

The probability $h(n)$ that a request for object $n$ can be satisfied by an LRU cache of size $C$ is
\begin{equation}
h(n) = 1 - e^{-q(n)t_c}
\label{eq:hn}
\end{equation}
where the ``characteristic time'' $t_c$ is the unique solution to the equation 
\begin{equation}
C = \sum_{n=1}^N h(n).
\label{eq:identity}
\end{equation}
The overall hit rate is $\theta=\sum q(n)h(n)/\sum q(n)$. Note that performance depends on the relative values of the $q(n)$ and not their absolute values: under the IRM, hit rates do not depend on traffic intensity. It is usual to order the $q(n)$ in decreasing order (object 1 is the most popular, object $N$ the least) but the approximation does not depend on this.  

The approximation extends to objects of variable size. Let object $n$ be of size $s_n$ and assume $s_n \ll C$ for all $n$. The hit rate is still given by (\ref{eq:hn}) where $t_c$ now solves 
\begin{equation}
C = \sum_1^N h(n)s_n.
\label{eq:identitysize}
\end{equation}

Suppose $s_n$ is measured in chunks of constant size and that objects are downloaded as a sequence of chunks. First assume all chunks are always requested whenever the object is requested so that each chunk inherits the object's popularity. The per-object hit rate is still given by (\ref{eq:hn}) and (\ref{eq:identitysize}) and is clearly identical to the hit rate for each of the object's chunks. Note, however, that equations (\ref{eq:hn}) and (\ref{eq:identitysize}) would also apply had we assumed the IRM applied to chunks, ignoring therefore the obvious correlation between successive requests for chunks of the same object. We conclude that, under condition $s_n \ll C$ for $1 \le n \le N$, the hit rate can be accurately evaluated assuming the IRM also applies to chunks.

Assume now users do not necessarily request every chunk of an object, as is common when the objects in question are streamed videos, for instance. The popularity law per chunk will be different, $q'(m)$ say for $1\le m \le \sum_n s_n$. However, under the condition  $s_n \ll C$ we expect the chunk hit rates to still be well-approximated as above by assuming the IRM at chunk level. Moreover, under quite severe user impatience, we observe empirically in Section \ref{sec:impatience} below that the popularity law under impatience is practically the same (after reordering and to within a multiplicative constant) as the per-chunk law assuming no impatience. The Che approximation thus applies even when users are impatient and the overall hit rate $\theta$ is largely independent of the characteristics of this impatience.

\subsection{A three part popularity law}
While we believe the IRM is a valid model for our purposes, it remains very difficult in general to estimate the popularity law $q(n)$. For example, it is not possible to directly infer the instantaneous popularity of a given object from a measurement of the number of requests for that object over a period of one week, say. While request rates can be assumed constant, as in the IRM, for a short periods, it is clear that an object's popularity can vary significantly over periods of hours and days.  Fortunately, one type of content does allow an estimation of its popularity law. This is the set of torrents advertised by a BitTorrent search engine like \emph{mininova.org}. 

Dan and Carlsson and co-authors have analyzed a large data set obtained for torrents referenced by \emph{mininova} \cite{DC10,CDMA12}. They kindly provided us with some of their raw data that we have used to derive a per-chunk popularity law.

A first data set allows us to classify some $2.9 \times 10^6$ active torrents according to the number of leechers they had at the time of capture, namely 8pm GMT on 15 Sept. 2008. We retained $1.6 \times 10^6$ torrents that had at least one leecher (the others were active because they had at least one seed). The number of leechers is a measure of instantaneous popularity but does not immediately give the required request rates. By Little's law, the request rate $q(n)$ for torrent $n$ is the expected number of leechers divided by the expected download time. Estimating the former by the measured number of leechers $l(n)$ and assuming the latter is proportional to the torrent size in chunks $s_n$, we have $q(n) \propto l(n)/s_n$.

Unfortunately, the leecher file does not include torrent size data. However, a second data set provides torrent size for another set of torrents, as described in \cite{CDMA12}. The torrents in both data sets are identified by the usual content hash and we were able to match leechers and size for some 330~000 of the $1.6 \times 10^6$ torrents. We filled in the blanks by preserving as far as possible the size distribution of the torrents of known size for every fixed number of leechers.

\begin{figure}[tp]
 \centering
 \includegraphics[scale=.8]{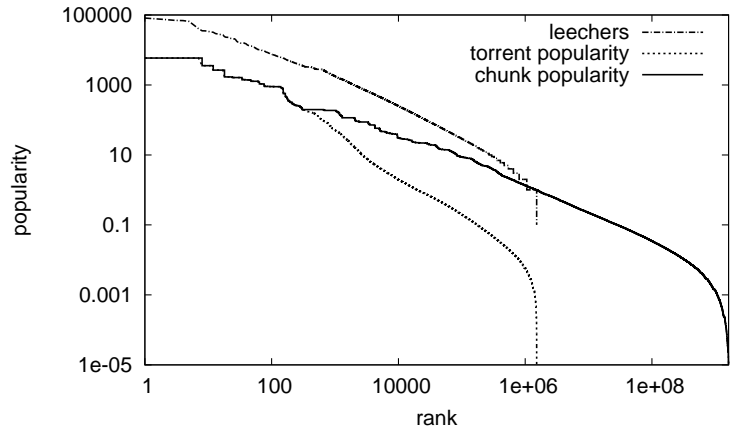}
 \caption{Popularity law for torrents; number of leechers $l(n)$, revised torrent popularity $q(n)$, derived chunk popularity}
 \label{fig:popularity}
\end{figure}

Figure \ref{fig:popularity} shows three popularity plots. The top one is the original leechers against rank plot and is identical to the corresponding curve in Figure 2 of \cite{DC10}. The bottom curve is normalized torrent popularity $l(n)/s_n$ plotted against rank. Finally, the middle curve plots the popularity of 1 MB chunks against rank. It is derived by stretching the second curve by counting $s_n$ equal popularity chunks for torrent $n$ for $1 \le n \le N$. The average torrent size is around 1 GB so the total number of chunks is $1.6 \times 10^9$ for a total of 1.6 PB.  Note that the last two curves coincide for the 300 most popular torrents since these are all just 1 chunk in size (the actual size is rounded up to 1 MB for cache occupancy though the intensity $l(n)/s_n$ is derived using the actual value of $s_n$ in bytes).

\begin{figure}[tp]
 \centering
 \includegraphics[scale=.8]{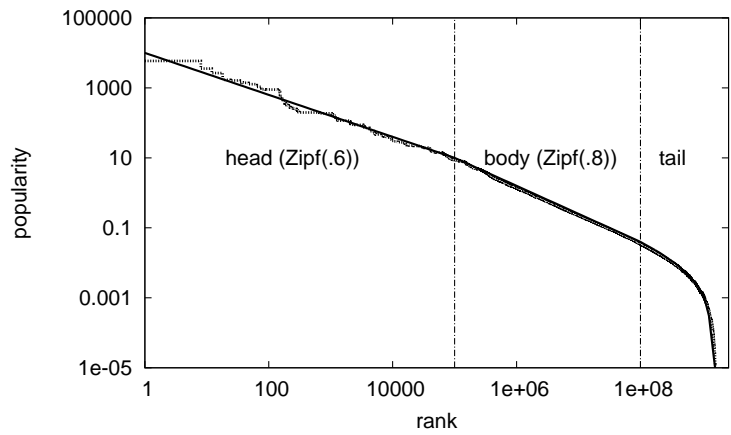}
 \caption{Popularity law for torrents; fit }
 \label{fig:popularityfit}
\end{figure}

We use the per-chunk popularity law below for our evaluations of the memory-bandwidth tradeoff. More precisely, we fit a sequence of power law segments to represent this law, as shown in Figure \ref{fig:popularityfit}. This law has three main components that we label ``head'', ``body'' and ``tail''. The head is flat and roughly parallel to a power law $1/n^{.6}$. In the following, we refer to such a power law with exponent $\alpha$ as Zipf($\alpha$). The body is somewhat steeper and parallel to Zipf(.8). Lastly, popularity drops rapidly in the tail and we fit it by a sequence of Zipf segments with exponent increasing from 1 to 15.  

Given the number of chunks in each segment and their individual popularities, we can apportion download traffic as follows: the head represents roughly 10\% (for $10^5$ chunks), the body 59\% ($10^8$ chunks) and the tail 31\% ($1.5\times 10^9$ chunks). Note the significant volume contributed by the tail despite the very low popularity of the torrents from which it is composed.

\subsection{Hit rate as function of cache size}
\label{sec:hitvcache}

\begin{figure}[tp]
 \centering
 \includegraphics[scale=.8]{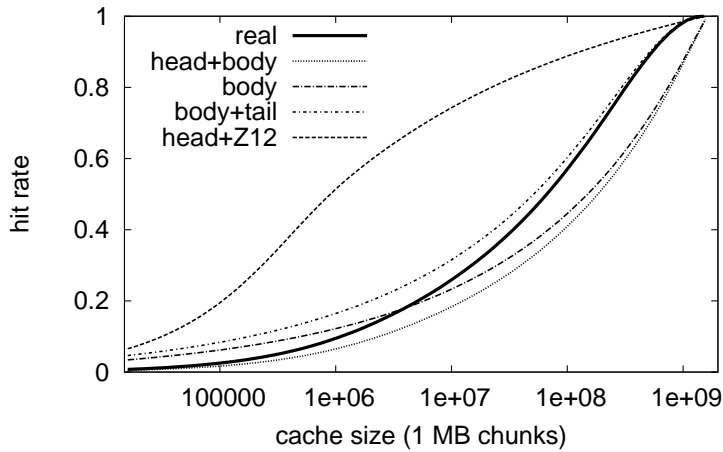}
 \caption{Hit rate as function of cache size for different popularity laws: real from Figure \ref{fig:popularityfit}, head + body, body alone, body + tail, head + Zipf(1.2) body.}
 \label{fig:Danpop}
\end{figure}

We evaluate the sensitivity of hit rate estimates to the degree of approximation in representing the popularity law. Numerical results derived using the Che approximation are presented in Figure \ref{fig:Danpop}. The hit rate for the ``real'' 3-component law of Figure \ref{fig:popularityfit} is given by the bold line. The simple expedient of assuming popularity is Zipf(.8) for $1 \le n \le 1.6\times 10^9$ is inaccurate for both small and large caches. The other two curves show how the head and tail are important to accurately predict hit rates of small and large caches, respectively (when head or tail is missing, the body is extended to ranks 1 or $1.6\times 10^9$, respectively). 

Lastly, we recall the impact of supposing the popularity is more accentuated than measurements suggest. Specifically, we suppose $q(n)$ follows the head up to $n=10^5$ and then decreases rapidly as Zipf(1.2).  It is clear from Figure \ref{fig:Danpop} that such an assumption would lead to widely inaccurate hit rate estimations.

 \begin{figure}[h]
                \centering
                        \fontsize{12}{12}\selectfont 
                \resizebox{!}{7.2cm}{\input{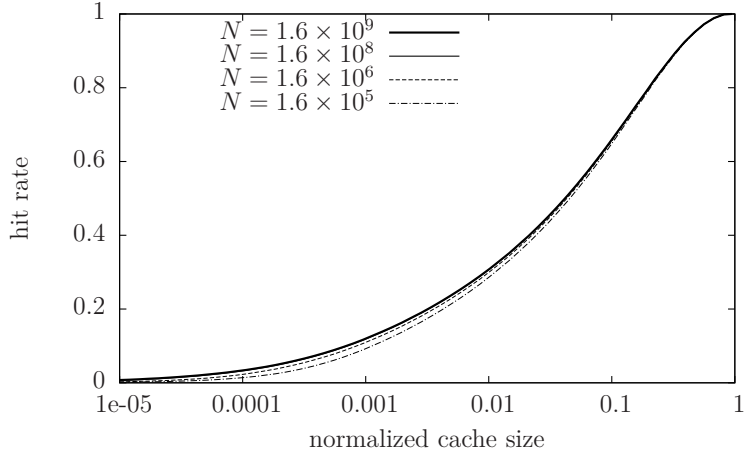}}
               \setlength{\belowcaptionskip}{-3pt}
                \caption{Hit rate as function of $C/N$ for different values of $N$ using the empirical popularity law. }
                \label{fig:convergence}
 \end{figure}%

Consider now the impact of the catalogue size. We assume the shape of the popularity law is retained while scaling the rank in proportion to catalogue size. We refer below to this scaled function as the ``empirical popularity law'', for whatever catalogue size $N$ is appropriate. For example, a catalogue of $N=1.6\times 10^6$ chunks has components delimited by $n=100$ and $n=10^5$ with the same slopes on the loglog plots. It is known that for Zipf($\alpha$) popularity with $\alpha<1$, the hit rate for cache size $C$ and catalogue $N$ tends to a limit function $\theta(C/N)$ as $N\to \infty$ \cite{Fill96}. Unsurprisingly, the hit rate for the empirical law of Figure \ref{fig:popularityfit} has the same behaviour, as illustrated in Figure \ref{fig:convergence}. The hit rate $\theta$ expressed as a function of $C/N$ is practically the same for $N>10^6$. 

\subsection{Impact of decreasing chunk popularities}
\label{sec:impatience}
In the absence of more precise real data, we use the popularity law of Figure \ref{fig:popularityfit} as if it were universal for all content. One possible objection is that, while torrents are useful only when their download is complete, other forms of content retrieval suffer from user impatience so that it is not correct to assume all chunks of the same object have equal popularity. To evaluate the impact of users interrupting a streaming video, say, we take the empirical data of Figure \ref{fig:popularity} and modify it as follows.

We retain torrents of size $s$ satisfying 10 MB $\le s \le$ 1 GB, both to limit the data set volume and because this range would be typical of video content. For each torrent with $l$ leechers and size $s$, we assume the popularity of chunks decreases linearly from $l/s$ to $0.3\times l/s$ implying that only 30\% of objects are downloaded to the end. The chunks are then resorted in order of decreasing popularity. 

\begin{figure}[tp]
 \centering
 \includegraphics[scale=.8]{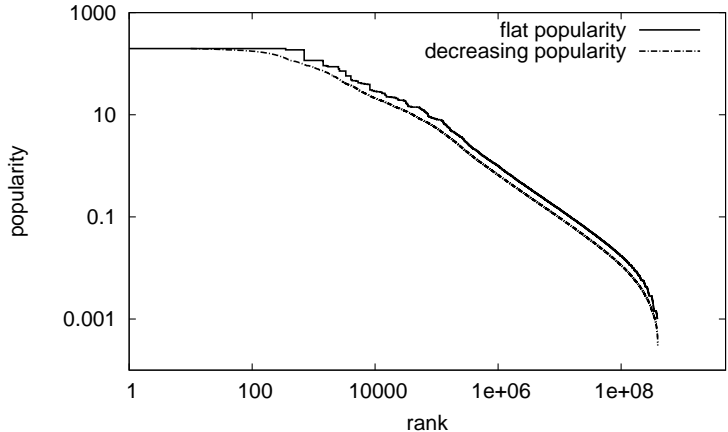}
 \caption{Popularity law of chunks assuming either all chunks inherit object popularity (flat) or chunk popularity decreases linearly to 30\% of initial value}
 \label{fig:impatience}
\end{figure}

Figure \ref{fig:impatience} compares the popularity laws of this model and that of the original where all chunks have popularity $l/s$. Except for the most popular chunks on the left, the impact of impatience is to reduce popularity by a factor of about 1/3 while preserving the slope of the law. Since hit rates in the IRM are determined by the relative values of the $q(n)$, we conclude that the hit rate is hardly changed by the assumed downloader impatience (see Sec. \ref{sec:Che}). This is so because the difference in popularity between distinct objects largely outweighs the difference in popularity between chunks of the same object.

In the following we use the derived empirical popularity law as if it applies to all content and not just torrents. We might vary the catalogue size but we assume the shape of the law, its head, body and tail, remain the same. The justification is that this law is derived from the best popularity measurements we have and that no published measurements for other types of content suggest the shape would be radically different. Popularity against rank has consistently been shown to exhibit  power law behaviour for the body with an exponent less than 1 and not too different to .8. Moreover, the petabyte catalogue size is also representative of other types of content like web pages or user-generated content \cite{FRRS12}.

\section{A two-level cache hierarchy}
\label{sec:hierarchy}

        \begin{figure}[h]
                       \centering
                \includegraphics[scale=.8]{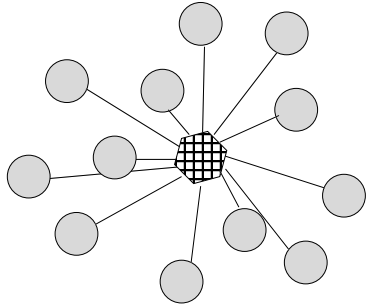}
               \caption{Two-level cache hierarchy with $S=12$ level-1 sites and 1 level-2 site}
                \label{fig:hierarchy}
        \end{figure}%

We use the Che approximation and the empirical popularity law to quantify the memory-bandwidth tradeoff for a simple, symmetrical 2-level cache hierarchy (Fig. \ref{fig:hierarchy}). Level-1 caches serve distinct sets of local users and have identical demand characteristics. If a requested chunk is absent at the local level-1 cache, the request is forwarded to the single level-2 cache. Replacement is LRU at each level. Level-1 caches occupy $S$ sites. They have capacity for $C$ chunks while the level-2 cache has capacity $\bar{C}$. Users generate a total busy period download traffic of $T$ bit/sec. The entire content catalogue consists of $N$ chunks with the empirical popularity law of Figure \ref{fig:popularityfit}.

\subsection{Cost difference}
To evaluate the memory-bandwidth tradeoff we consider the cost of the considered network as a function of $C$, excluding fixed costs. We refer to this as the \emph{cost difference}. In particular, we fix $S$, and therefore ignore the cost of the access network, and fix an overall network hit rate target  $\Theta$, and therefore ignore the cost of retrieving content from beyond the considered network. Fixing $\Theta$ implies $\bar{C}$ is a function of $C$ and can be calculated using the Che approximation, on assuming cache occupancies are independent \cite{FRRS12}. Denote the level-1 hit rate by $\theta$. It is a function of $C$ and $N$ but not of $T$.

We assume bandwidth costs are proportional to the traffic flowing between level-2 and level-1 with a constant of proportionality $k_b$. Bandwidth thus costs $T(1-\theta)k_b$. 

We assume the cost of caching is due to two factors: the cost of memory, supposed proportional to capacity, and the cost of serving content, supposed proportional to peak demand in bit/sec. Denoting the constants of proportionality by $k_m$ and $k_s$, respectively, total caching costs are $(S C + \bar{C}) k_m + T\left(1 + (1-\theta)\right) k_s$. The last  term sums all traffic served by level-1 and the fraction  $(1-\theta)$ also served by level-2.

Excluding the constant service cost $Tk_s$, the memory-bandwidth tradeoff is characterized by the cost difference,
\begin{equation}
\Delta(C) = T \left(1-\theta\right) (k_b+k_s) + (S C + \bar{C}) k_m. 
\label{eq:Delta}
\end{equation}

\subsection{Cost estimates}
\label{sec:costs}

To progress we need plausible estimates for the constants $k_b$, $k_m$ and $k_s$. 

We set $k_b= \$15$ per Mbps as a nominal monthly charge for bandwidth. There are no publicly available references to justify this choice. It is derived from some leased-line pricing reported on the web and some data given privately by an operator. Factor $k_b$ is intended to cover the cost of both transport and routers. 

It is noteworthy that the price of bandwidth is decreasing quite rapidly. For instance, the blog DrPeering\footnote{http://drpeering.net/white-papers/Internet-Transit-Pricing-Historical-And-Projected.php} reports that the average monthly price per Mbps for IP transit was only \$5 in 2010 and declining fast. The Ethernet alliance states the price of Ethernet bandwidth needs to decrease at an annual rate of 20\% to meet demand forecasts\footnote{http://www.networkworld.com/news/tech/2012/041012-ethernet-alliance-258118.html?page=1}.

We set the nominal monthly unit cost of memory to $k_m=\$0.15$ per gigabyte. This estimate is derived from advertised storage costs from cloud providers like Amazon and is meant to cover all costs of storage using whatever devices are appropriate. 

This cost appears to be decreasing more rapidly than that of bandwidth. Cost trends have been tracked between 1957 to 2012 by J. C. McCallum\footnote{http://www.jcmit.com/memoryprice.htm} revealing a regular decrease rate of 40\% per annum. 
 
We estimate the cost $k_s$ of serving content to be negligible relative to $k_b$. This observation derives from cloud service download charges of around \$.10 per GB or roughly \$.10 per Mbps peak rate. In other words, in the evaluations below, we assume $k_b+k_s \approx k_b = \$15$ per Mbps per month.

\subsection{Evaluation}

 \begin{figure}[h]
                \centering
                \fontsize{12}{12}\selectfont 
                \resizebox{!}{7.5cm}{\input{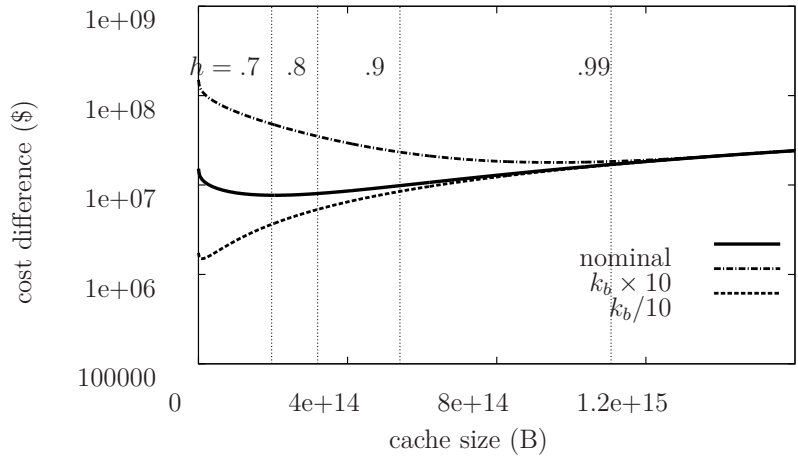}}
                \setlength{\abovecaptionskip}{-9pt}
                \caption{Cost difference (\$) against cache size (bytes) for the data of Section \ref{sec:costs} (nominal), for bandwidth cost $k_b$ multiplied by 10, and for $k_b$ divided by 10; overall hit rate $\Theta$ is 100\%. }
                \label{fig:deltacost}
 \end{figure}%

\begin{figure}[h]
        \centering
        \fontsize{12}{12}\selectfont 
                \resizebox{!}{7cm}{\input{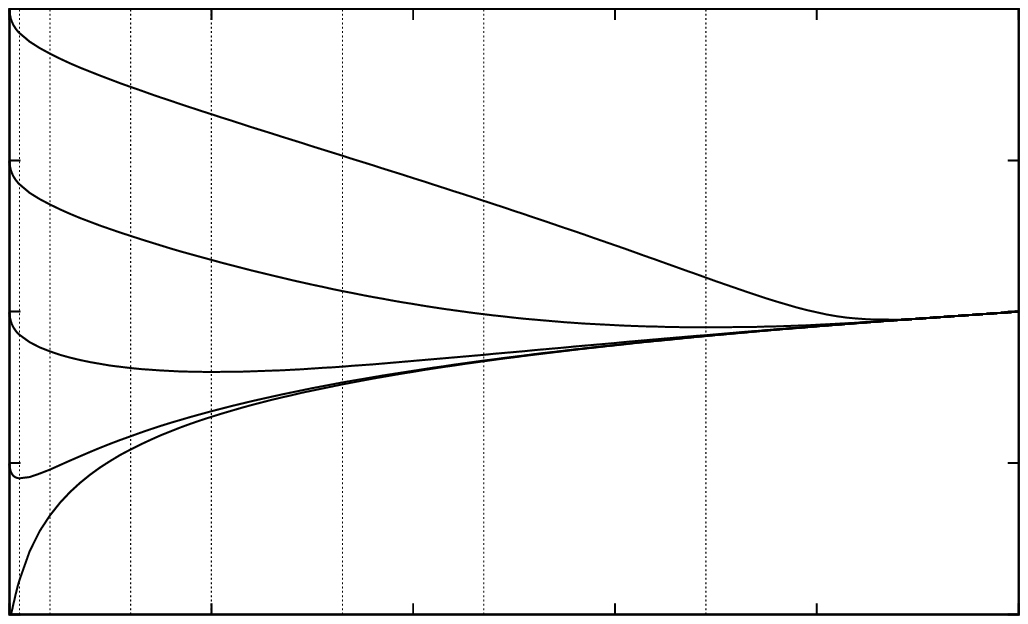}}
		\setlength{\abovecaptionskip}{-9pt}
               \caption{Normalized cost difference $\delta(c)$ against normailized cache size $c$ for empirical popularity law: five curves for each graph correspond to different values of $\Gamma$ $(= \delta(0))$ ranging from .01 to 100; vertical lines indicate level-1 hit rates.  $\Theta=1$, b/w cost $\propto$ traffic.}
                \label{fig:basedelta}
        \end{figure}%
      
\begin{figure}[h]
        \centering
        \fontsize{12}{12}\selectfont 
                \resizebox{!}{7cm}{\input{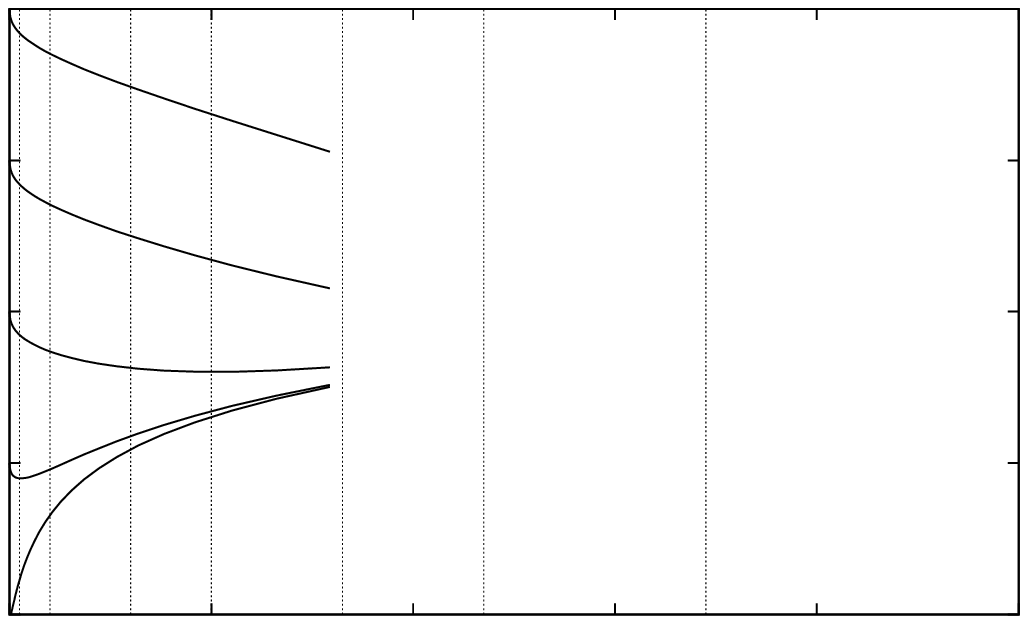}}
		\setlength{\abovecaptionskip}{-9pt}
               \caption{Normalized cost difference $\delta(c)$ against normailized cache size $c$ for empirical popularity law: five curves for each graph correspond to different values of $\Gamma$ $(= \delta(0))$ ranging from .01 to 100; vertical lines indicate level-1 hit rates.  $\Theta=0.9$, b/w cost $\propto$ traffic.}
                \label{fig:hit90delta}
        \end{figure}%
        
\begin{figure}[h]
        \centering
        \fontsize{12}{12}\selectfont 

		\resizebox{!}{7cm}{\input{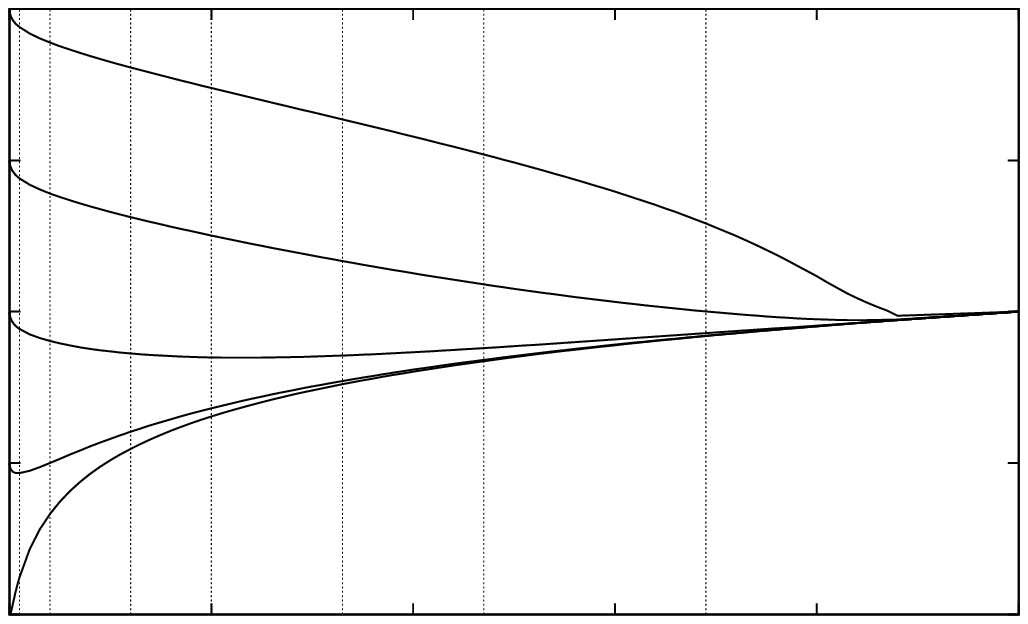}}
		 \setlength{\abovecaptionskip}{-9pt}
                \caption{Normalized cost difference $\delta(c)$ against normailized cache size $c$ for empirical popularity law: five curves for each graph correspond to different values of $\Gamma$ $(= \delta(0))$ ranging from .01 to 100; vertical lines indicate level-1 hit rates. $\Theta=1$, b/w cost $\propto$ (traffic)$^{.75}$.}
                \label{fig:scaledelta}
        \end{figure}

Figure \ref{fig:deltacost} plots $\Delta$ as a function of $C$ with parameter values $T=1$ Tbps, $S=100$, and $\bar{C} \equiv N$, corresponding to an overall hit rate $\Theta=1$. These data are intended to be representative of a national ISP network in a country like France. We use the empirical popularity law with a total catalogue size $N=1.6\times 10^9$ chunks of 1 MB. The impact of varying these parameters is considered later.

The middle curve is plotted using the constants $k_b$ and $k_m$ given in Section \ref{sec:costs}. The maximum monthly cost of bandwidth (for $C=0$) is \$15M while the maximum cost of caching (for $C=N$) is \$24M. Cost is minimized at \$7.5M for $C=200$ TB with a level-1 hit rate of 80\%. The upper and lower curves show the impact of an order of magnitude error in the estimated cost factor $k_b$. The shapes of the curves would be the same had we instead made the same order of magnitude errors in $k_m$. 

The curves demonstrate that there is limited scope for optimization by splitting required cache capacity between levels 1 and 2. For most estimates of $k_b$ and $k_m$ it is either best to cache (nearly) all content at level-1 or none at all. The choice depends on the relative values of the maximum cost of memory ($C=N$) and the maximum cost of bandwidth ($C=0$). The cost gain in the nominal data case is 50\% of the maximum cost of bandwidth.  
To more fully explore the tradeoff, we consider a normalized cost comparison in the next section.

\subsection{Normalized costs}

For the sake of simplicity, we again suppose the overall hit rate objective is $\Theta=1$ implying $\bar{C}\equiv N $. We can then remove the term in $\bar{C}$ from the cost difference (\ref{eq:Delta}). We also suppress $k_s$, supposed negligible compared to $k_b$. Denote by $\Gamma$ the ratio of the maximum cost of bandwidth to the maximum cost of memory at level-1,
$$\Gamma=\frac{Tk_b}{SNk_m}.$$
From the results of Section \ref{sec:hitvcache} (cf. Fig. \ref{fig:convergence}), $\theta$ for given $C$ and $N$ is actually a function $\theta(c)$ of the normalized cache size $c=C/N$. We can therefore quantify the tradeoff in terms of a normalized cost difference,
\begin{equation}
\delta(c) = \Gamma (1-\theta(c)) + c,
\label{eq:littledelta}
\end{equation}
where units are normalized such that $\delta$ varies between $\Gamma$ and 1 as $c$ increases from 0 to 1.

Figure \ref{fig:normalizedcost} shows how $\delta(c)$ depends on different parameters. Figure \ref{fig:basedelta} generalizes Figure \ref{fig:deltacost}, demonstrating how the optimal level-1 cache size increases from 0 to nearly the entire catalogue as $\Gamma$ increases. As $\Gamma$ encapsulates all the model parameters it is possible to make the following general observations:
\begin{itemize}
\item since the number of sites $S$ determines the position of caches in the access, aggregation or core network, the relative values of $T$ and $N$ determine the network position at which it is most cost-effective to cache content,
\item  if the catalogue $N$ were smaller (if, say, it relates only to some limited, identifiable set of VoD content), it tends to be optimal to cache (nearly) all chunks at level-1, even for large $S$,
\item as costs evolve (e.g., $k_b$ decreases by 20\% per year, $k_m$ decreases by $40\%$) while traffic $T$ increases (e.g., by 40\%), the trend is for $\Gamma$ to increase quite quickly (e.g., increasing 10 fold in less than 4 years),
\item if $\Gamma>10$, it is not useful for cost reasons to cache any content at level-2,
\item it is unlikely to be worthwhile to use level-1 caches that can hold just a small fraction of the catalogue.
\end{itemize}

Figure \ref{fig:hit90delta} shows the effect of limiting the overall hit rate objective to 90\% (we then have $\bar{C} < N$ but this discrepancy has negligible impact on costs). It reveals that the cost difference can be derived in practice by truncating the curves of Figure \ref{fig:basedelta} such that $\theta(c)\le .9$.  If $\Gamma\ge 10$, for example, it makes no sense to limit the hit rate to a value less than 99\%, even without considering the cost of content retrieval from outside the considered network.

For Figure \ref{fig:scaledelta} we suppose bandwidth costs are sub-linear in offered traffic as this may  more realistically reflect scale economies. Specifically, and somewhat arbitrarily, we suppose costs are proportional to traffic to the power .75 (cf. \cite{NPSS95}) and set $\Gamma = T^{.75}k_b/(SNk_m)$. $\Gamma$ is the ratio of maximum bandwidth cost to maximum memory cost for this model. Results show that scale economies reinforce our observation that optimal level-1 cache sizing is typically ``(nearly) all or nothing''.

It is never optimal to cache the entire catalogue at level-1 because of the significant tail of the empirical popularity law. Had we ignored this and assumed a Zipf(.8) law for body and tail, an  ``all or nothing'' rule of thumb would be more appropriate. If the body of the popularity law were Zipf($\alpha$) with $\alpha > 1$, on the other hand, our numerical results (not shown here) suggest the scope for optimization would be more pronounced. We recall, however, that all published measurements of content popularity confirm that $\alpha>1$ would be an unrealistic assumption.

\section{Alternative cache networks}
\label{sec:alternatives}

We consider how the observations of the last section for a two-level hierarchy generalize to alternative cache networks.

\subsection{Hierarchy with more than two levels}
We have not explicitly evaluated a hierarchical cache network with more than 2 levels, agreeing with Borst and co-authors that ``it rarely pays off to install caches at more than one or two levels'' \cite{BGW10}. It is necessary, however, to choose the right level in the network for placing the caches. This determines the number of sites $S$ and the respective network costs above and below these sites. The results in Section \ref{sec:hierarchy} together with a cost estimate of access networks would allow an appraisal of the best choice. 

\subsection{Load sharing}

        \begin{figure}[h]
                \centering
                \includegraphics[scale=.8]{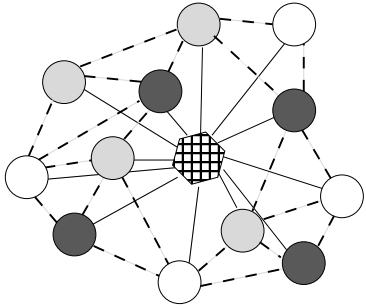}
               \caption{Mesh of level-1 caches with $S=12$ sites and $P=3$ content partitions depicted by the node shading}
                \label{fig:mesh}
        \end{figure}%

When cache size is limited, it is desirable to aggregate their capacity by load sharing. For instance, the proposal by Li and Simon \cite{LS11} is to divide the chunk catalogue into $P$ partitions and assign these to neighbouring level-1 caches. Users address their requests to the nearest cache responsible for the partition the chunk belongs to. The partition would typically be determined from a hash of the chunk name and a simple routing protocol would designate the nearest cache. 

Figure \ref{fig:mesh} illustrates this idea using dashed lines to represent paths between level-1 caches. Cache misses at level-1 overflow to the level-2 cache as before. We assume partitioning preserves symmetry: all caches manage the same size catalogue $N/P$ and receive identical demand. Let the level-1 cache size be $C/P$, other notation being as in Section \ref{sec:hierarchy}.

First observe that the level-1 hit rate is still $\theta(c)$ where $c=C/N$. This follows from the convergence results depicted in Figure \ref{fig:convergence}. To evaluate the normalized cost difference $\delta_{\textsc{ls}}(c)$ for this network we introduce a new factor $k'_b$ for the unit cost of inter-cache bandwidth at level-1. It is straightforward to show that
\begin{equation}
\delta_{\textsc{ls}}(c) = \delta(c) + (1-\frac{1}{P})(\Gamma \frac{k'_b}{k_b}-c)
\label{eq:deltaLS}
\end{equation}
where $\delta(c)$ is given by (\ref{eq:littledelta}).

This formula shows that no partitioning ($P$=1) is preferable if $\Gamma k'_b/k_b> c$. In general, it is necessary to account for the dependence of $k'_b$ on $P$ and the optimal choice is difficult to characterize. However, the cost data and trends discussed in Section \ref{sec:hierarchy}  suggest partitioning would be of limited utility in practice unless there is an imposed capacity limit. For example, if $\Gamma \ge 10$ it is not worth using partitions unless $k'_b < k_b/10$.

\subsection{Cooperative caching}

\begin{figure}[h]
       \centering
        \fontsize{12}{12}\selectfont 

                   \resizebox{!}{7cm}{\input{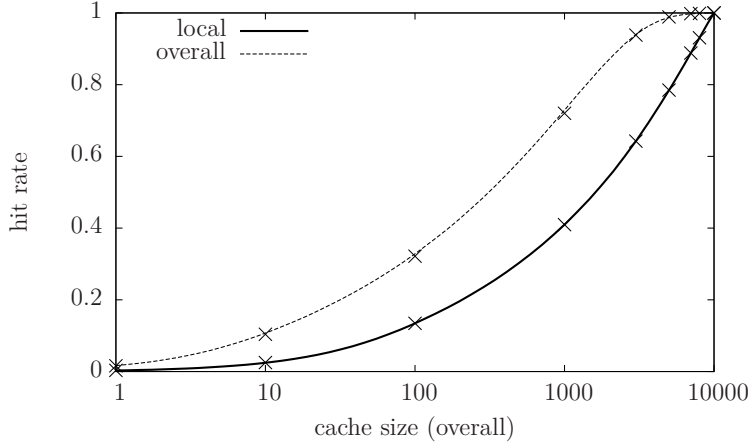}}
		\setlength{\abovecaptionskip}{-9pt}
               \caption{Hit rates for cooperative cache networks. Simulation v analysis: $N=10000$, Zipf(.8) popularity, $P=10$}
                \label{fig:simulation}
        \end{figure}
        
\begin{figure}[h]
       \centering
        \fontsize{12}{12}\selectfont 
                \resizebox{!}{7cm}{\input{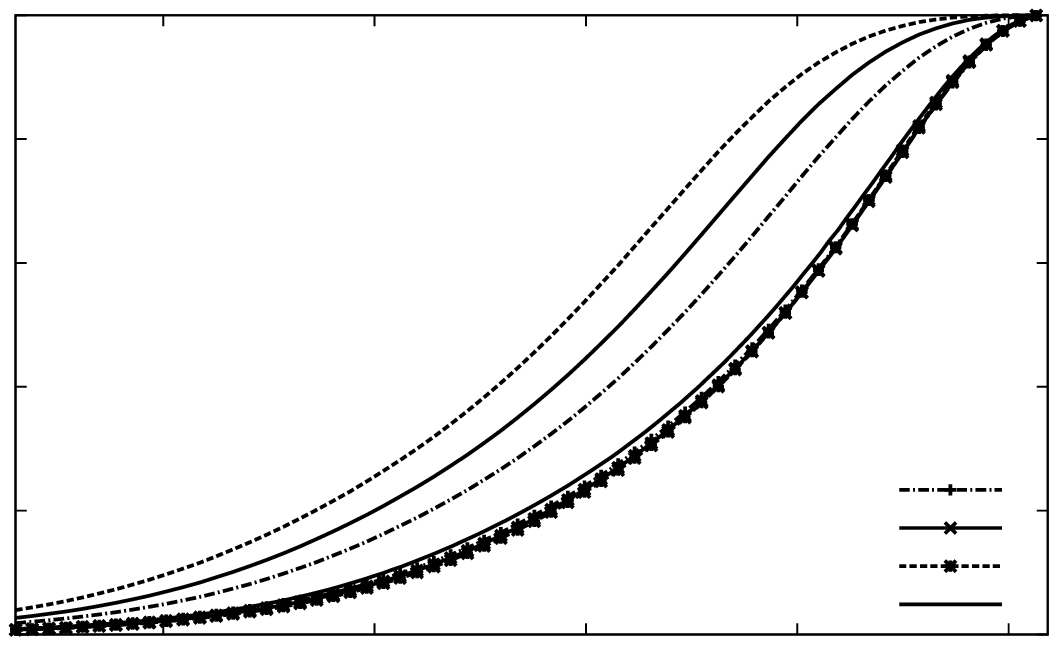}}
		\setlength{\abovecaptionskip}{-9pt}
               \caption{Hit rates for cooperative cache networks. Overall and local hit rates,  $\theta$ and $\theta'$, against cache capacity $C$}
                \label{fig:localcap}
        \end{figure}%

\begin{figure}[h]
       \centering
        \fontsize{12}{12}\selectfont 
		\resizebox{!}{7cm}{\input{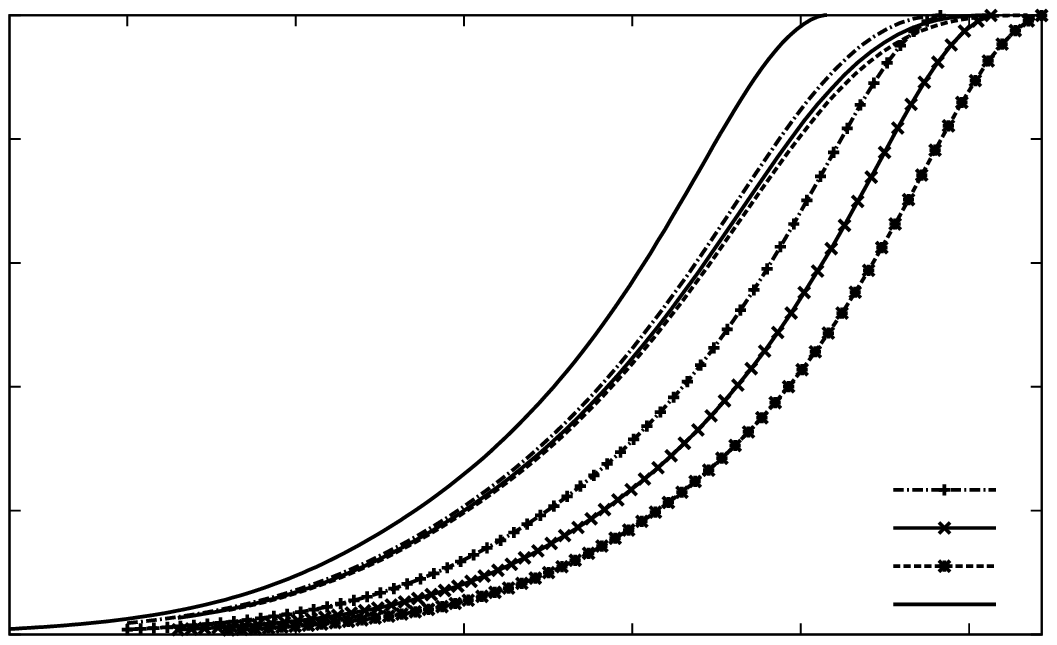}}
		 \setlength{\abovecaptionskip}{-9pt}
                \caption{Hit rates for cooperative cache networks. Overall and local hit rates,  $\theta$ and $\theta'$, against overall capacity $PC$}
                \label{fig:overallcap}
        \end{figure}

A possible cooperative caching scheme proposed by Ni and Tsang \cite{NT05} builds on load sharing. Users first address requests to their local cache. If this is unsuccessful, the request is forwarded to a cache designated for the partition to which the requested chunk belongs (except if the local cache already belongs to this partition). If this ``interaid'' fails requests go to the level-2 cache as before. This scheme allows simple request routing, like load sharing, while avoiding longer paths for the most popular chunks. For simplicity, we again assume partitioning and routing preserve symmetry.

To calculate hit rates we adapt the Che approximation (cf. Sec. \ref{sec:Che}).  The popularity of any chunk $n$ that belongs to a cache's designated partition increases to $q'(n)=q(n)(1+(P-1)e^{-q(n)t_c})$. The equation to determine the characteristic time $t_c$ is modified as follows: 
$$ \sum_{n=1}^N \frac{1}{P}(1-e^{-q'(n)t_c)} + (1-\frac{1}{P}) (1-e^{-q(n)t_c}) = C.$$
Given $t_c$ and assuming cache occupancies are statistically independent, we can calculate both the hit rate at the local cache, denoted $\theta'$, and the overall level-1 hit rate, $\theta$:
\setlength{\multlinegap}{0mm}
\begin{multline*}
\theta' = \sum_{n=1}^N q(n) \left(\frac{1}{P}(1-e^{-q'(n)t_c}) + (1-\frac{1}{P}) (1-e^{-q(n)t_c})\right) / \sum_{n=1}^N q(n),
 \end{multline*}
$$\theta= \theta' + \sum_{n=1}^N q(n) (1-\frac{1}{P}) e^{-q(n)t_c}(1-e^{-q'(n)t_c})/ \sum_{n=1}^N q(n).$$

We have tested the accuracy of this approximation by simulation for a particular case: 10000 chunks with Zipf(.8) popularity are divided into 10 partitions. Results shown in Figure \ref{fig:simulation} demonstrate remarkable accuracy. We ran the simulation long enough that the results do not change when the run time is multiplied by 10. Since the accuracy of the Che approximation increases with system size (cf. \cite{FRR12}), we are confident the hit rates derived below for a large catalogue of $1.6 \times 10^9$ chunks are very accurate.

Figures \ref{fig:localcap} and \ref{fig:overallcap} plot $\theta'$ and $\theta$ against local cache size $C$ and overall cache capacity $PC$, respectively,  for 4 values of $P$ including 1 (corresponding to the 2-level hierarchy of Section \ref{sec:hierarchy}). Note first from Figure \ref{fig:localcap} that $\theta'$ as a function of $C$ is practically independent of $P$. Level-1 interaid significantly increases the request rate of some chunks but since these are distributed uniformly over the catalogue, the \emph{shape} of the popularity law plotted against rank hardly changes. This is similar to the lack of impact of decreasing chunk popularity observed in Section \ref{sec:impatience} (Fig. \ref{fig:impatience}).

Figure \ref{fig:overallcap} shows that $\theta$ as a function of $PC$ converges rapidly as $P$ increases. The limit $\theta(PC,N)$ is somewhat lower than the hit rate $\theta(C,N)$ obtained without partitions ($P$=1). Results show the considered cooperative scheme requires (about 3 times) more memory for the same hit rate $\theta$ as a single cache in the 2-level hierarchy with $S/P$ sites. 

To compare this with previous approaches we suppose caches are sized so that the level-1 hit rate $\theta$ is the same for all. Let the cache size for coordinated caching be  $\tilde{C}$. Let $\tilde{h}$ be the hit rate at the local cache for chunks that do not belong to its own partition. Using cost factors $k_m$, $k_b$ and $k'_b$ introduced above, the normalized cost difference for this scheme can be written,
$$ \delta_{\textsc{cc}}(c) = \frac{\tilde{C}}{N} + \Gamma (1-\theta(c)) + \Gamma \frac{k'_b}{k_b} (1-\frac{1}{P})(1-\tilde{\theta}(\tilde{C},N)).$$
From the numerical results of Figure \ref{fig:cooperation} we have $\tilde{C} > C/P$. Moreover, since $\theta'(\tilde{C},N) \approx h(C/N)$ and $\tilde{\theta} \lesssim \theta'$, since the foreign chunks are less popular, we have $\tilde{\theta}(\tilde{C}, N) \lesssim \tilde{\theta}(c/P) \lesssim \theta(c/P)$ where $\lesssim$ means ``less than'' in the sense of the approximate convergence of hit rates in Figure \ref{fig:convergence}. These inequalities allow us to deduce,
\begin{equation}
\delta_{\textsc{cc}}(c) \gtrsim \delta(c) + (1-\frac{1}{P}) \left(\Gamma \frac{k'_b}{k_b} (1-\theta(c/P))-c\right).
\label{eq:deltaCC}
\end{equation}

As for load sharing, it is impossible to make a definitive statement about the value of this type of cooperation. However, if we are correct to believe the cost of bandwidth increasingly outweighs that of memory, the simple duplication of large caches at a suitably chosen level in the network is unlikely to be far from optimal for the future Internet.

\section{Conclusions}
The ICN memory-bandwidth tradeoff depends crucially on the hit rate realized by a cache of given capacity. To evaluate this we have used the Che approximation with a realistic traffic model derived from measurements of the popularity law of BitTorrent content retrieval. We showed that this approximation remains accurate when applied to chunks, despite the correlated request process and even accounting for decreasing chunk popularity due to frequent incomplete downloads.

Our analysis is based on the performance of symmetric cache networks assuming linear dependence of cost on capacity. These simplifications reveal quite robust structural properties that are unlikely to be disproved by more detailed models. We note also that such refinements would still be very hard to correctly parameterize. 

Under our best guess cost and traffic assumptions, a 2-level cache hierarchy realizing the optimal tradeoff would equip level-1 caches to capture around 70\% of download traffic with a capacity equivalent to 10\% of the entire catalogue (i.e., caches of around 100 TB). However, accounting for cost and traffic trends, it will soon (within 4 years, say) be optimal to achieve a hit rate of 99\% by caching up to 75\% of the catalogue. 

The above figures apply to a choice of traffic and network parameters that is meant to be representative of a country like France. More general understanding is derived from normalized cost formulas where the critical quantity is shown to be the ratio of the total cost of bandwidth without caching to the total cost of storing the entire catalogue in each level-1 cache. The normalized tradeoff formulas can be used to determine the optimal siting of caches for given costs and traffic volumes. We can, for instance, evaluate the advantage of isolating popular content (e.g., a particular VoD catalogue) to be cached very close to end-users, while storing the petabytes of general content (web, UGC, file sharing) somewhere much closer to the network core.  

Cooperation between level-1 caches may lead to cost reduction but this again depends on relative costs of memory and bandwidth. We have considered two cooperative strategies and derived cost formulas that characterize the tradeoff. Under our best assumptions about unit costs and their evolution, the formulas suggest cooperation brings little cost advantage over the simple cache hierarchy.

Our deduction from the above is that the future Internet is less likely to be a network of content store augmented routers than a loosely interconnected network of local data centers. Since these data centers should be equipped to cater for the large majority of content downloads, and these count for the majority of Internet demand, traffic circulating above them would be reduced by an order of magnitude. A further advantage is that a data center would be better able to perform the necessary higher level functions of a content distribution network.  

\section*{Acknowledgment}
This work was partially funded by French ANR project CONNECT under grant ANR-10-VERS-001.

\bibliographystyle{abbrv}
\bibliography{cache}

\end{document}